\documentclass[a4paper,11pt]{article}
\usepackage{amssymb}
\usepackage{graphicx}
\usepackage{amsmath}
\usepackage{indentfirst}
\usepackage{mathrsfs}
\usepackage{cite}
\usepackage{hyperref}

\begin{document}

\title{K\"{a}ll\'{e}n-Lehmann Representation of Noncommutative Quantum
Electrodynamics}
\author{R. Bufalo$^{1}$\thanks{%
rbufalo@ift.unesp.br}~, T.R. Cardoso$^{1}$\thanks{%
cardoso@ift.unesp.br}~, B.M. Pimentel$^{1}$\thanks{%
pimentel@ift.unesp.br}~ \\
\textit{{$^{1}${\small Instituto de F\'{\i}sica Te\'orica (IFT/UNESP), UNESP
- S\~ao Paulo State University}}} \\
\textit{\small Rua Dr. Bento Teobaldo Ferraz 271, Bloco II Barra Funda, CEP
01140-070 S\~ao Paulo, SP, Brazil}\\
}
\maketitle
\date{}

\begin{abstract}
Noncommutative (NC) quantum field theory is subject of many analyses
on formal and general aspects looking for deviations and, therefore,
potential noncommutative spacetime effects. Within of this large class,
we may now pay some attention to the quantization of NC field theory on
lower dimensions and look closely at the issue of dynamical mass
generation to the gauge field. This work encompasses the
quantization of the two-dimensional massive quantum electrodynamics and
three-dimensional topologically massive quantum electrodynamics. We begin by
addressing the problem on a general dimensionality, making use of the
perturbative Seiberg-Witten map to, thus, construct a general action, to
only then specify the problem to two and three dimensions. The quantization
takes place through the K\"{a}ll\'{e}n-Lehmann spectral representation and
Yang-Feldman-K\"{a}ll\'{e}n formulation, where we calculate the respective spectral density
function to the gauge field. Furthermore, regarding the photon two-point function, we discuss how its infrared behavior is related to the
term generated by quantum corrections in two dimensions, and, moreover, in three dimensions,
we study the issue of nontrivial $\theta$ dependent corrections to the
dynamical mass generation.
\end{abstract}



\section{Introduction}

The quantization of spacetime, in particular a quantum fuzzy spacetime,
comes as a natural consequence to support the coexistence of quantum field
theory and gravity, demanding, thus, a radical change in our concepts of
geometry \cite{14}. Noncommutative (NC) spacetime characterized by
canonical Heisenberg-like (commutation relations) Moyal brackets
\begin{equation}
\left[ x^{\mu },x^{\nu }\right] _{\star}\equiv x^{\mu }\star x^{\nu }-x^{\nu }\star x^{\mu }=i\theta ^{\mu \nu },
\label{eq 0.0}
\end{equation}%
and its associated $\star $ product (Weyl symbols)%
\begin{equation}
\phi \left( x\right) \star \phi \left( x\right) =\phi \left( x\right) \exp
\left( \frac{i}{2}\overleftarrow{\partial}_{\mu } \theta ^{\mu \nu }\overrightarrow{\partial} _{\nu
}\right) \phi \left( x\right) ,  \label{eq 0.1}
\end{equation}%
have received much attention over the course of the last two decades. Moreover, since $\theta^{\mu\nu}$ is a constant antisymmetric object,
theories defined on such noncommutative spacetime are considered to violate Lorentz invariance, a subtle, but highly important issue that
has been richly debated \cite{23,13}. Actually,
noncommutative spatial coordinates have a well-known realization in physics:
the quantized confined motion of particles in a constant magnetic field,
sufficiently strong so that the projection on the lowest Landau level can be
justified, is described by noncommutative coordinates. Within the context of
field theory, it was only after the work of Seiberg and Witten \cite{1} that
the NC gauge theories have acquired a prominent place in several
discussions. Many of the intriguing fundamental phenomena regarding NC gauge theories
are results from the nonlocality of their interactions, for instance, UV/IR mixing
problem \cite{15}, loss of unitarity \cite{16}, and violation of Lorentz
symmetry \cite{17}. Nevertheless, investigation on these
features has pointed out, for instance, that a suitable definition of
time-ordering operation restores the unitarity in a NC field theory \cite{18}
and many other interesting issues \cite{42}.

A successful and extensively studied approach to a perturbative analysis of (%
\textit{nonlocal}) NC field theory is obtained if one exploits the
Seiberg-Witten (SW) map \cite{1,41}. Namely, they have shown that a NC $\star $%
-gauge theory should be a gauge equivalent to an ordinary counterpart defined
on a commutative spacetime, i.e., the SW transformation maps a $%
\star $-gauge invariant NC expression into a gauge invariant ordinary
expression. In such approach, one can study systematically in a
perturbative way the effects of noncommutativity in a \textit{local} quantum
field framework. Furthermore, a different route may be taken in obtaining the aforementioned SW map without referring to
string theory. This consists of letting the theory be an
enveloping algebra valued one, that one, thus, can write the SW map
of NC fields for arbitrary non-Abelian gauge groups, such
as $SU(N)$ \cite{40}. Therefore, it is clear that the SW map of NC fields presents itself as an interesting tool in the understanding
of the physical predictions and also to check the behavior of the NC theory itself, such as renormalizability.

Based on the above discussion, it is interesting to know how the
noncommutativity affects established properties of conventional theories,
i.e., the study of NC extensions of well-studied quantum field theories and
to look then for NC effects on its deviations, since it is generally found
that such extensions behave in interesting and nontrivial ways.
Therefore, for this purpose, we will investigate here the NC effects first
on the celebrated massive Schwinger model \cite{2,3} [quantum electrodynamics ($QED$) in
two-dimensions] and next on the $QED$ in three dimensions, Abelian gauge
theory, Maxwell-Chern-Simons, or simply Maxwell electrodynamics coupled to
fermions \cite{5,6,21,30,24,36}. One may say that the main goal of physics is to explain phenomena in nature, and perhaps
even to explain why physical nature dwells in four dimensions; however, the means that we have
come to employ in reaching this goal so far are sufficiently
intricate that it has proven useful to wander into lower-dimensional worlds, with the wishful thought
that in a simpler setting we can learn useful things about the well-recognized four-dimensional problems.
Both classes of field theory have been
extensively investigated along the years, and have long been recognized as
laboratories where important theoretical ideas, such as infrared problems, dynamical
mass generation, and confinement, to name few, were tested in a simple
setting, especially on condensed matter systems, e.g., the quantum Hall effect. Actually, some NC properties of them
were discussed previously in Refs.\cite{4,4a} on massless $QED_{2}$ and in Refs.\cite%
{8,7} on $QED_{3} $.

To analyze the quantized theory, we will make use of a general framework,
possessing a solid physical basis which is the spectral representation of K%
\"{a}ll\'{e}n-Lehmann \cite{9}. Furthermore, in order to accomplish that, we
will make use of the Yang-Feldman-K\"{a}ll\'{e}n formalism \cite{10} which is based on
the Heisenberg operators equations of motion and asymptotic conditions.
Also, extensions of these formalisms were proposed and studied in different scenarios: a noncommutative
spacetime \cite{11} and also for a Lorentz-violating field theory \cite{22}.

In this paper, we revisit $QED_{2}$ and $QED_{3}$ embedded in a
noncommutative spacetime from the point of view of K\"{a}ll\'{e}n-Lehmann
formalism, in particular, their radiative corrections at one loop and the \textit{possible}
higher-order term \cite{4a} and dynamically mass generation, respectively. We start by discussing the general properties and symmetries of
NCQED in a generic dimension $\omega $ in Sec.\ref{sec:1}. Next, we
discuss and make use of the Seiberg-Witten map to determine the action. Next, we present the expressions
to the Lagrangian functions where we will perform the analysis; both expressions are obtained by
taking into account the dimensionality of spacetime. From that, we derive
the Euler-Lagrange equations of motion, and from the Yang-Feldman-K\"{a}ll\'{e}n formalism,
we construct the solution for these equations. Moreover, another
important feature of Yang-Feldman-K\"{a}ll\'{e}n formalism, is that one works exclusively
with the free asymptotic fields so that all quantities are mathematically
well defined. In Sec.\ref{sec:2} we review and discuss, in general terms,
the K\"{a}ll\'{e}n-Lehmann representation, and obtain the spectral density
function to the gauge field. In Sec.\ref{sec:3}, we present
the calculation and results, at two-particle contribution, to the spectral density function of gauge
field, both calculated on two and three dimensions. In
particular, based on a previous result \cite{4a}, we discuss in detail the correct expression for the
self-interaction contribution of the gauge field spectral function in the
two-dimensional case, which originates from the infrared sector. In Sec.\ref%
{sec:4} we summarize the results, and present our final remarks and
prospects.

\section{Seiberg-Witten Map on NCQED}

\label{sec:1}

Before addressing our problem, \textit{per se}, we start this section by reviewing the
quantum electrodynamics of massive fermions in an $\omega $-dimensional
noncommutative Minkowski spacetime. As a consequence of the nontriviality
of the star product, the ordinary theory acquires a non-Abelian-like structure,
namely, the NCQED model action is defined in the following way \footnote{%
According to our notation, "hatted" quantities will represent objects
defined on the noncommutative spacetime, and "unhatted" ones will stand as
defined on the ordinary spacetime.}:%
\begin{equation}
\mathcal{A}=\int d^{\omega }x\left[ -\frac{1}{4}\hat{F}_{\mu \nu }\star \hat{%
F}^{\mu \nu }+\hat{\bar{\psi}}\star \left( i\gamma ^{\mu }\hat{D}_{\mu }\hat{%
\psi}-m\hat{\psi}\right) \right] ,  \label{eq 1.0}
\end{equation}%
where the field strength and covariant derivative are defined as%
\begin{eqnarray}
\hat{F}_{\mu \nu } &=&\partial _{\mu }\hat{A}_{\nu }-\partial _{\nu }\hat{A}%
_{\mu }-ig\left[ \hat{A}_{\mu },\hat{A}_{\nu }\right] _{\star },  \notag \\
\hat{D}_{\mu }\hat{\psi} &=&\partial _{\mu }\hat{\psi}-ig\hat{A}_{\mu }\star
\hat{\psi}.  \label{eq 1.1}
\end{eqnarray}%
The action is invariant under the finite noncommutative $U\left( 1\right) $ $%
\star $-gauge transformations:%
\begin{eqnarray}
\hat{\psi}^{\prime }\left( x\right) &=&U\left( x\right) \star \hat{\psi}%
\left( x\right) ,  \notag \\
\hat{A}_{\mu }^{\prime }\left( x\right) &=&\frac{i}{g}U\left( x\right) \star
\hat{D}_{\mu }U^{-1}\left( x\right) ;  \label{eq 1.2}
\end{eqnarray}%
or, rather, in its infinitesimal form,%
\begin{eqnarray}
\hat{\delta} _{\hat{\sigma} }\hat{\psi} &=&i\hat{\sigma}\left( x\right) \star \hat{\psi}%
\left( x\right) ,  \notag \\
\hat{\delta} _{\hat{\sigma} }\hat{A}_{\mu } &=&\frac{1}{g}\partial _{\mu }\hat{\sigma}%
\left( x\right) -i\left[ \hat{A}_{\mu },\hat{\sigma}\right] _{\star }.
\label{eq 1.3}
\end{eqnarray}%
where $U\left( x\right) =\left(e^{i\sigma(x)}\right)_{\star}$ is defined by an infinite
series of multiple $star$ products of scalar function $\hat{\sigma}\left( x\right) $.

Given the properties of the Moyal product, the product of two functions
(distributions) is integrated, giving the same result as the ordinary product of
functions. This implies that in this framework, the free propagator
expressions do not change; i.e., they have the same expression as in the
commutative (ordinary) theory. Therefore, in the case of covariant
perturbative analysis, the signal of noncommutativity is encompassed on the
theory's vertices. However, we are interested in studying effects of
noncommutativity in a perturbative way by making use of the SW map \cite{1}, which allows one to write an NC $\star $-gauge theory as a
gauge theory in ordinary spacetime.

The SW map of NC fields in a gauge invariant theory can be derived from a
gauge equivalence relation as it stands for the gauge field and parameter
\begin{eqnarray}
\hat{\delta} _{\hat{\sigma} }\hat{A}_{\mu}\left(A;\theta\right)&=&\hat{A}_{\mu}\left(A+\delta_{\sigma}A;\theta\right)-\hat{A}_{\mu}\left(A;\theta\right)\notag\\
&=& \delta _{\sigma }\hat{A}_{\mu}\left(A;\theta\right)\label{eq 1.4d}\\
\hat{\delta} _{\hat{\sigma} }\hat{\sigma}\left(\sigma,A;\theta\right)&=& \delta _{\sigma }\hat{\sigma}\left(\sigma,A;\theta\right) \label{eq 1.4b}
\end{eqnarray}
as for the matter field
\begin{eqnarray}
\hat{\delta} _{\hat{\sigma} }\hat{\psi}\left(\psi,A;\theta\right)= \delta _{\sigma }\hat{\psi}\left(\psi,A;\theta\right) \label{eq 1.4c}
\end{eqnarray}
in such a way that the NC fields are functionals of the ordinary fields: $A_{\mu }$, $\psi $, and $\sigma $ are the ordinary gauge field, fermion
field, and gauge transformation parameter, respectively. To the lowest nontrivial order in $\theta $,
one finds the solution to the SW map, \footnote{ Because of the freedom in the solutions,
ambiguities are present on different solutions of the maps and are reflected in different
coefficients obtained in each of the solutions constructed. For the interested
reader, we refer to Ref.\cite{41} for a complete disclosure of the general features of the SW map.}%
\begin{eqnarray}
\hat{A}_{\mu } &=&A_{\mu }-\frac{g}{2}\theta ^{\lambda \sigma }A_{\lambda
}\left( 2\partial _{\sigma }A_{\mu }-\partial _{\mu }A_{\sigma }\right)
+O\left( \theta ^{2}\right) , \\
\hat{\psi} &=&\psi -\frac{g}{2}\theta ^{\lambda \sigma }A_{\lambda }\partial
_{\sigma }\psi +O\left( \theta ^{2}\right) ,  \label{eq 1.4} \\
\hat{\sigma} &=&\sigma -\frac{g}{2}\theta ^{\lambda \sigma }A_{\lambda
}\partial _{\sigma }\sigma +O\left( \theta ^{2}\right) .
\end{eqnarray}%
Consequently, an important feature of this map is that it preserves gauge orbits, and so $\star $%
-gauge invariance is, therefore, rendered into an ordinary gauge invariance as contained in Eqs.\eqref{eq 1.4d}-\eqref{eq 1.4c}.
Utilizing this map, in terms of the ordinary quantities, we arrive at the following $O\left( \theta \right) $
modified form for the NC quantum electrodynamics action \cite{43}:
\begin{eqnarray}
\mathcal{A} &=&\int d^{\omega }x\bigg[-\frac{1}{4} \left( 1-\frac{g}{2}%
\left( \theta .F\right) \right) F_{\mu \nu }F^{\mu \nu } -\frac{g}{2}\theta ^{\lambda
\sigma }F_{\mu \lambda }F_{\nu \sigma }F^{\mu \nu } +\left( 1-\frac{g}{4}\left( \theta .F\right) \right) \bar{\psi}i\gamma
.D\psi\notag \\
&& +\frac{g}{2}\theta ^{\lambda \sigma }\bar{\psi}i\gamma ^{\mu
}F_{\lambda \mu }D_{\sigma }\psi -m\left( 1-\frac{g}{4}\left( \theta
.F\right) \right) \bar{\psi}\psi \bigg].  \label{eq 1.5}
\end{eqnarray}
An interesting modification seen on the gauge sector at leading order in $%
\theta $ is the presence of self-interaction terms. Although the NC gauge
sector resembles the Yang-Mills theories, it is the noncommutative structure
of spacetime which causes the nonlinearity of the field strength in the
gauge connection. Furthermore, for $\omega >2$, the action \eqref{eq 1.5}
provides a suitable framework on the investigation of a Lorentz-violating
extension of $QED$, once all the $\theta $-dependent terms violate Lorentz
symmetry \cite{23}.

It should be emphasized that the only needed ingredient in our development in the K\"{a}ll\'{e}n-Lehmann
representation is the translational invariance, \footnote{The Lorentz invariance is only required in a way in deriving the
dependence and tensor structure of the spectral function with the momentum $p$ as it will be shown next.} which is always satisfied
in the noncommutative theory. Nevertheless,
the Lorentz symmetry (rotations and boosts) is preserved only if $\theta^{\mu\nu}$
transforms as a tensor \cite{13}, taking different constant values in different frames.
Moreover, it also follows that the Seiberg-Witten map has an explicit Lorentz-invariant form provided that $\theta$ transforms like a Lorentz tensor,
in accordance with the previous discussion \cite{13}.


\subsection{($ 1+1$) and ($ 2+1$) NCQED}

At this point, we have considered a field theory defined on an $\omega $%
-dimensional spacetime. Looking for simplifications of the $\theta $ terms
of action \eqref{eq 1.5}, we shall consider separately two particular cases:
$\left( 1+1\right) $- and $\left(
2+1\right) $-dimensional spacetime. Thus, taking into account the
dimensionality of spacetime, one arrives at%
\begin{eqnarray}
\mathcal{L}_{1+1}=\bar{\psi}\left( i\gamma . D-m\right) \psi +%
\frac{mg}{4}\left( \theta .F\right) \bar{\psi}\psi -\frac{1}{4}\left( 1+%
\frac{g}{2}\left( \theta .F\right) \right) F_{\mu \nu }F^{\mu \nu }-\frac{1}{%
2\xi }\left( \partial _{\mu }A^{\mu }\right) ^{2}, \label{eq 1.6}
\end{eqnarray}%
for a two-dimensional spacetime, whereas, for a three dimensions, one gets%
\begin{eqnarray}
\mathcal{L}_{2+1}=\bar{\psi}\left( i\gamma .D -m\right) \psi +%
\frac{mg}{4}\left( \theta .F\right) \bar{\psi}\psi  -\frac{\mu}{
4}\epsilon ^{\sigma \nu \lambda }A_{\sigma }F_{\nu \lambda }-\frac{1}{4}\left( 1+%
\frac{g}{2}\left( \theta .F\right) \right) F_{\mu \nu }F^{\mu \nu }-\frac{1}{2\xi }%
\left( \partial _{\mu }A^{\mu }\right) ^{2}, \label{eq 1.7}
\end{eqnarray}%
where in both Lagrangian functions, a gauge-fixing term on the Lorenz condition was
inserted. Moreover, on the three-dimensional Lagrangian
\eqref{eq 1.7}, $\epsilon ^{\mu \nu \lambda }$ is the totally antisymmetric
Levi-Civit\`{a} tensor, $\mu$ denotes the coupling of the topological term, and we
have also made use of the NC extension of the Chern-Simons action derived in
Ref.\cite{7}, where it was showed that under the SW map, the NC Chern-Simons
theory reduces to its commutative counterpart to all orders of $\theta $.

Once we have developed the models of interest, i.e., obtained the Lagrangian functions for $\left( 1+1\right) $ and
$\left( 2+1\right) $ dimensions, Eqs.\eqref{eq
1.6} and \eqref{eq 1.7}, respectively, we are ready to proceed with our
development. Since we aim to discuss both theories on the framework of K\"{a}%
ll\'{e}n-Lehmann representation \cite{9}, we have to calculate next the equations
of motion of the fields and subsequently find their solution in terms of
the Yang-Feldman-K\"{a}ll\'{e}n equations for Heisenberg operators \cite{10}.

On the fermion sector, we obtain the same equation of motion for both cases,%
\begin{equation}
\left( i\gamma ^{\mu }D_{\mu }-m\right) \psi =-\frac{mg}{4}\left( \theta
.F\right) \psi ;  \label{eq 1.8}
\end{equation}%
whereas for the gauge sector, we obtain from Eq.\eqref{eq 1.6},%
\begin{eqnarray}
&&\partial _{\alpha }F^{\alpha \beta }+\frac{1}{\xi }\partial ^{\beta
}\partial _{\lambda }A^{\lambda }+g\bar{\psi}\gamma ^{\beta }\psi   +\frac{g}{4}\partial _{\alpha }\bigg[
\theta ^{\alpha \beta }F_{\mu \nu }F^{\mu \nu }+2\left( \theta .F\right)
F^{\alpha \beta }-2m\theta ^{\alpha \beta }\bar{\psi}\psi \bigg]  =0,\label{eq 1.9}
\end{eqnarray}%
and from Eq.\eqref{eq 1.7}, it yields
\begin{eqnarray}
&&\partial _{\alpha }F^{\alpha \beta }+\frac{1}{\xi }\partial ^{\beta
}\partial _{\lambda }A^{\lambda }-\frac{\mu}{2}\epsilon ^{\alpha \mu \beta
}F_{\alpha \mu }+g\bar{\psi}\gamma ^{\beta}\psi +\frac{g}{4}\partial _{\alpha }\bigg[ \theta ^{\alpha \beta
}F_{\mu \nu }F^{\mu \nu }+2\left( \theta .F\right) F^{\alpha \beta
}-2m\theta ^{\alpha \beta }\bar{\psi}\psi \bigg]  =0. \label{eq 1.10}
\end{eqnarray}
In accordance with the general statement of the Yang-Feldman-K\"{a}ll\'{e}n formulation, with
suitable boundary conditions at $t=\pm \infty $, we find, from Eq.\eqref{eq 1.8} that the Heisenberg operators $\psi $ and $\bar{\psi}$
satisfy the equations \cite{12}
\begin{eqnarray}
\psi _{A}\left( x\right) &=&\psi _{A}^{in}\left( x\right) -\int d^{\omega
}yS_{_{AB}}^{ret}\left( x-y\right) \eta _{B}\left( y\right) ,  \notag \\
\bar{\psi}_{A}\left( x\right) &=&\bar{\psi}_{A}^{in}\left( x\right) -\int
d^{\omega }y\zeta _{B}\left( y\right) S_{_{AB}}^{adv}\left( y-x\right) ,
\label{eq 1.11}
\end{eqnarray}%
where the currents are given by%
\begin{equation}
\eta =g\gamma .A\psi -\frac{mg}{4}\left( \theta .F\right) \psi ,\quad \zeta
=g\bar{\psi}\gamma .A-\frac{mg}{4}\left( \theta .F\right) \bar{\psi}.
\label{eq 1.12}
\end{equation}%
The retarded (advanced) fermionic Green's function is%
\begin{equation}
S^{ret\left( adv\right) }\left( x\right) =\frac{1}{\left( 2\pi \right)
^{\omega }}\int d^{\omega }p\frac{1}{\gamma .p-m\mp i\epsilon p_{0}}e^{-ipx}.
\label{eq 1.13}
\end{equation}%
In such formulation the asymptotic $in$- and $out$-\emph{fields} (operator-valued
fields) satisfy free-field equations, and, thus, can be decomposed into
positive and negative frequency parts \cite{12}. Furthermore, the
Yang-Feldman-K\"{a}ll\'{e}n equation for the gauge field solution of Eqs.\eqref{eq 1.9} and \eqref{eq 1.10}%
takes the form%
\begin{equation}
A_{\mu }\left( x\right) =A_{\mu }^{in}\left( x\right) -\int d^{\omega
}y\Delta _{\mu \sigma }^{ret}\left( x-y\right) j^{\sigma }\left( y\right) ,
\label{eq 1.14}
\end{equation}%
where%
\begin{eqnarray}
j^{\beta }=g\bar{\psi}\gamma ^{\beta }\psi  +\frac{g}{4}\partial _{\alpha }%
\bigg[ \theta ^{\alpha \beta }F_{\mu \nu }F^{\mu \nu }+2\left( \theta
.F\right) F^{\alpha \beta }-2m\theta ^{\alpha \beta }\bar{\psi}\psi \bigg] .\label{eq 1.15} 
\end{eqnarray}%
The first term in the current Eq.\eqref{eq 1.15} is the usual $U\left( 1\right) $ gauge interaction term, the
second and third terms are related with the photon self-interaction, whereas
the fourth term is a Yukawa interaction type. Aside from the integral
dimensionality, the difference between the two- and. three-dimensional
solutions are the Green's functions. That for $\left( 1+1\right) $ dimensions reads%
\begin{equation}
\Delta _{\mu \sigma }^{ret\left( adv\right) }\left( x\right) =\int \frac{d^{2}h}{
\left( 2\pi \right) ^{2}}\left[ \eta _{\mu \sigma }-\frac{h_{\mu
}h_{\sigma }}{h^{2}}\right] \frac{e^{-ihx}}{h^{2}\mp i\epsilon h_{0}},
\label{eq 1.16}
\end{equation}%
whereas for $\left( 2+1\right) $ dimensions, it reads
\begin{eqnarray}
\Delta _{\lambda \nu }^{ret\left( adv\right) }\left( x\right) =\int \frac{d^{3}h}{\left(
2\pi \right) ^{3}}\left[ \eta _{\lambda \nu }-\frac{h_{\lambda }h_{\nu }}{%
h^{2}}-\frac{i\mu}{h^{2}}\epsilon _{\lambda \nu \alpha }h^{\alpha }\right] \frac{e^{-ihx}
}{h^{2}-\mu^{2}\mp i\epsilon h_{0}}.  \label{eq 1.17}
\end{eqnarray}%
As we have succeeded in deriving the Lagrangian densities for two- and three-dimensional spacetime, and from
them found the equations of motion and, subsequently, the Yang-Feldman-K\"{a}ll\'{e}n
solution for the (operator-valued) fields $A_{\mu }$, $\psi $, and $\bar{\psi%
}$, we are now in position to calculate the spectral density functions for
the contributions of one- and two-particles for the gauge
field. But before such calculation, we will briefly review the K\"{a}ll\'{e}n-Lehmann
spectral representation.


\section{K\"{a}ll\'{e}n-Lehmann representation: Exact propagators}

\label{sec:2}

Once we have obtained the Yang-Feldman-K\"{a}ll\'{e}n equations on the ordinary spacetime,
Eqs. \eqref{eq 1.11} and \eqref{eq 1.14}, the next step consists (different from the dispersion relations approach to noncommutative models
\cite{11}) of the investigation of the spectral functions following the
ordinary K\"{a}ll\'{e}n-Lehmann representation \cite{9,12}. However, before that
calculation, let us write some lines and describe the general prescription
of the K\"{a}ll\'{e}n-Lehmann representation. Let $A_{\mu }\left( x\right) $ be a
vector field on the Heisenberg representation. \footnote{%
Here we will maintain ourselves a discussion of vector fields embedded in a
four-dimensional spacetime, but, the derivation for other fields and
dimensionality is rather direct.} The vacuum expectation value of the
product of two fields at different points can be expressed as%
\begin{equation}
\left\langle \Omega \left\vert A_{\mu }\left( x\right) A_{\nu }\left(
y\right) \right\vert \Omega \right\rangle =\underset{n}{\sum }\left\langle
\Omega \left\vert A_{\mu }\left( x\right) \right\vert n\right\rangle
\left\langle n\left\vert A_{\nu }\left( y\right) \right\vert \Omega
\right\rangle ,  \label{eq 2.0}
\end{equation}%
where the completeness relation of the physical spectrum has been used $%
\left\{ |n\rangle \right\} $, i.e., $p_{\mu }|n\rangle =p_{\mu }^{\left(
n\right) }|n\rangle $, and $n$ represents all quantum numbers specifying a
state. Based only on general arguments about invariance and the spectrum of $%
p_{\mu }$, we will be able to determine the general expression of the exact
photon propagator. Furthermore, we use the translational invariance \cite{13} of the
theory $\left[ A_{\mu }\left( x\right) ,p_{\alpha }\right] =i\partial
_{\alpha }A_{\mu }\left( x\right) $, i.e., write $A_{\mu }\left( x\right)
=e^{ip\left( x-x_{0}\right) }A_{\mu }\left( x_{0}\right) e^{-ip\left(
x-x_{0}\right) }$, to obtain the following expression of the Wightman's function
\begin{equation}
\left\langle \Omega \left\vert A_{\mu }\left( x\right) A_{\nu }\left(
y\right) \right\vert \Omega \right\rangle =\int_{0}^{\infty }d\chi \rho
_{\mu \nu }\left( \chi \right) \Delta ^{\left( +\right) }\left( x-y;\chi
\right) ,  \label{eq 2.1}
\end{equation}%
where $\Delta ^{\left( +\right) }$ is the positive frequency part of the
Pauli-Jordan function. The theory's content, perturbative or
nonperturbative, is fully encoded on the spectral density function $\rho
_{\mu \nu }$. According to the Lorentz invariance, the dependence of $\rho $
on $p$ is only through $p^{2}$, in a such way that allows one to write%
\footnote{Here, $\tau $ stands for the step function.}%
\begin{eqnarray}
\rho _{\mu \nu }\left( q^{2}\right) \tau \left( q_{0}\right) =\left( 2\pi
\right) ^{3}\underset{n}{\sum }\delta ^{\left( 4\right) }\left( p^{\left(
n\right) }-q\right) \left\langle \Omega \left\vert A_{\mu }\left( 0\right)
\right\vert n\right\rangle \left\langle n\left\vert A_{\nu }\left( 0\right)
\right\vert \Omega \right\rangle .  \label{eq 2.2}
\end{eqnarray}%
This quantity is null for $q^{2}<0$, and it is real and non-negative for $%
q^{2}\geq 0$. In possess of the Wightman's function \eqref{eq 2.1}, one can
construct any propagator of its interest. For instance, it follows that
the exact Feynman propagator for the gauge field in the K\"{a}ll\'{e}n-Lehmann
spectral representation \eqref{eq 2.1} is given by%
\begin{equation}
i\mathcal{D}_{\mu \nu }\left( k^{2}\right) =\int_{0}^{\infty }d\chi \rho _{\mu \nu
}\left( \chi \right) \frac{1}{k^{2}-\chi -i\epsilon }.  \label{eq 2.3}
\end{equation}


\section{Spectral density function: perturbative example}

\label{sec:3}

After having obtained the Yang-Feldman-K\"{a}ll\'{e}n equations of the dynamical fields of
theory, at two and three dimensions, and subsequently, as a brief review, derived the main points of the spectral
density function for the gauge field and its relation
with the Feynman propagator as well, we are in position to perform an
explicit calculation for the one- and two-particle contributions. Moreover, we will make no distinction between the usual and
noncommutative contribution along our calculation, although it will be rather transparent in our resulting expressions,
in such a way that the reader can follow how the noncommutative contribution accounts to the usual one.
In order for unitarity to hold in the three-dimensional case, we will
assume that only $\theta _{ij}\neq 0$ \cite{16}, while the time coordinate
commutes with the space coordinates. \footnote{Therefore, noncommutative
quantum field theories in two-dimensional spacetime are
not unitary.}

\subsection{($1+1 $)-dimensional case}

On the K\"{a}ll\'{e}n-Lehmann formulation, the quantity to be initially
evaluated is the spectral density function \eqref{eq 2.2}. Moreover, because
of its tensor structure and Lorentz invariance, $\rho _{\mu \nu }$ can be
written as
\begin{equation}
\rho _{\mu \nu }\left( k^{2}\right) =\left( \eta _{\mu \nu }-\frac{k^{\mu
}k^{\nu }}{k^{2}}\right) \rho \left( k^{2}\right) ,  \label{eq 3.0}
\end{equation}%
where we have introduced the scalar spectral function $\rho $. The first
example here lies in evaluating the one-particle contribution,%
\begin{eqnarray}
&&\frac{\rho _{\mu \nu }\left( k^{2}\right) \tau \left( k_{0}\right) }{2\pi }%
=\int dp_{1}\delta ^{\left( 2\right) }\left( p_{1}-k\right)  \underset{j}{%
\sum }\left\langle \Omega \left\vert A_{\mu }\left( 0\right) \right\vert
p_{1},j\right\rangle \left\langle p_{1},j\left\vert A_{\nu }\left( 0\right)
\right\vert \Omega \right\rangle , \label{eq 3.1}
\end{eqnarray}
in which the one-particle state, a photon carrying momentum $p_{1}$ and
polarization $j$, is constructed such as $|p_{1},j\rangle =a_{j}^{\dag
}\left( p_{1}\right) |\Omega \rangle $. After some straightforward
calculation, it follows that the scalar function $\rho $ has the expression $
\rho ^{\left( 0\right) }\left( k^{2}\right) =2\delta \left( k^{2}\right)$.
Now, for the two-particle contribution, we have to be cautious and pay
attention to the theory's interaction structure. The general expression for
such contribution is given by
\begin{eqnarray}
\frac{\rho ^{\left( 1\right) }\left( k^{2}\right) \tau \left( k_{0}\right) }{%
2\pi }&=&\int dp_{1}dp_{2}\delta ^{\left( 2\right) }\left( k-p_{1}-p_{2}\right)  \underset{n,m}{%
\sum }\left\langle \Omega \left\vert A_{\mu }\left( 0\right) \right\vert
p_{1},n;p_{2},m\right\rangle \left\langle p_{1},n;p_{2},m\left\vert A^{\mu
}\left( 0\right) \right\vert \Omega \right\rangle , \notag \\
&=&\int dp_{1}dp_{2}\int d^{2}zd^{2}w\delta ^{\left( 2\right) }\left(
k-p_{1}-p_{2}\right) \eta _{\mu \nu }\Delta ^{\mu \sigma (ret)}\left(
-z\right) \Delta ^{\nu \rho (ret)}\left( -w\right)  \notag \\
&& \times\underset{n,m}{\sum }\left\langle \Omega \left\vert
j^{in}_{\sigma }\left( z\right) \right\vert
p_{1},n;p_{2},m\right\rangle\left\langle p_{1},n;p_{2},m\left\vert
j^{in}_{\rho }\left( w\right) \right\vert \Omega \right\rangle .
\label{eq 3.4}
\end{eqnarray}%
From the first to the second equality, we have made use of the Yang-Feldman-K\"{a}ll\'{e}n equation \eqref{eq 1.14},
where $\left( j^{in}\right) $ stands for the current \eqref{eq 1.15}
written in terms of the $in$-\emph{fields} and $\Delta _{\mu \sigma }^{ret}$ is
given by Eq.\eqref{eq 1.16}. Now, by analyzing the interaction structure of the
current \eqref{eq 1.15}, we see that it contains, besides the usual $QED$
term, a Yukawa kind of term and a photon self-interaction as well, information
that leads us to construct the two-particle state taking into account all
the intermediate interaction in the following form%
\begin{eqnarray}
\underset{n,m}{\sum }|p_{1},n;p_{2},m\rangle =\underset{i,j}{\sum }%
a_{i}^{\dag }\left( p_{1}\right) a_{j}^{\dag }\left( p_{2}\right) |\Omega
\rangle +\underset{r,s}{\sum }d_{r}^{\dag }\left( p_{1}\right) b_{s}^{\dag
}\left( p_{2}\right) |\Omega \rangle .  \label{eq 3.5}
\end{eqnarray}%
The first term is characterized by two photons carrying momenta and
polarization as $\left( p_{1},i\right) $ and $\left( p_{2},j\right) $,
whereas the second term corresponds to a fermion and anti-fermion pair,
characterized by their momenta and spin $\left( p_{1},r\right) $ and $%
\left( p_{2},s\right) $, respectively. Therefore, from the above discussion,
we find the following expression for the sum of the matrix elements:{\small
\begin{eqnarray}
\underset{n,m}{\sum }\left\langle \Omega \left\vert
j^{in}_{\sigma }\left( z\right) \right\vert
p_{1},n;p_{2},m\right\rangle\left\langle p_{1},n;p_{2},m\left\vert
j^{in}_{\rho }\left( w\right) \right\vert \Omega \right\rangle &=& \underset{r,s}{\sum }\left\langle \Omega \left\vert
j^{in}_{\sigma }\left( z\right) \right\vert
p_{1},r;p_{2},s\right\rangle \left\langle p_{1},r;p_{2},s\left\vert
j^{in}_{\rho }\left( w\right) \right\vert \Omega \right\rangle \notag \\
&&+\underset{i,j}{\sum }\left\langle \Omega \left\vert  j^{in}
_{\sigma }\left( z\right) \right\vert p_{1},i;p_{2},j\right\rangle
\left\langle p_{1},i;p_{2},j\left\vert  j^{in}_{\rho }\left(
w\right) \right\vert \Omega \right\rangle.  \notag \\\label{eq 3.6}
\end{eqnarray}%
}From the last result, we clearly see that we are left to calculate four
matrix elements: two of them are the usual contribution while the others
are related to the noncommutative contribution. As an example, we shall
evaluate the matrix elements related with the fermionic
interaction. Making use of the explicit expression for the current, for
instance, we have%
\begin{eqnarray}
\left\langle \Omega \left\vert  j^{in}_{\sigma }\left(
z\right) \right\vert p_{1},r;p_{2},s\right\rangle = g\left\langle \Omega
\left\vert \bar{\psi}^{in}\left( z\right) \gamma _{\sigma }\psi ^{in}\left(
z\right) \right\vert p_{1},r;p_{2},s\right\rangle -\frac{mg}{2}\left\langle
\Omega \left\vert \tilde{\partial}_{\sigma }\left[ \bar{\psi}^{in}\psi ^{in}%
\right] \left( z\right) \right\vert p_{1},r;p_{2},s\right\rangle ,
\label{eq 3.7}
\end{eqnarray}%
where we have introduced the notation $\tilde{a}^{\alpha }=a_{\mu }\theta
^{\mu \alpha }$. Now, using the free solutions of the (operator-valued)
fields \cite{12}, one obtains%
\begin{eqnarray}
\left\langle \Omega \left\vert  j^{in}_ {\sigma }\left(
z\right) \right\vert p_{1},r;p_{2},s\right\rangle = -\frac{mg}{\left( 2\pi
\right) }\sqrt{\frac{1}{E_{p_{1}}E_{p_{2}}}} e^{-iz\left( p_{1}+p_{2}\right) }\bar{v}\left( p_{1},r\right) \left[ \gamma _ {\sigma }+\frac{im}{2}\left( \tilde{p}_{1}+\tilde{p}_{2}\right)
_{\sigma }\right] u\left( p_{2},s\right) ,  \label{eq 3.8}
\end{eqnarray}%
with $p_{k}=\left( E_{p_{k}},\overrightarrow{p}_{k}\right) $ and $E_{p_{k}}=%
\sqrt{\overrightarrow{p}_{k}^{2}+m^{2}}$. The remaining matrix elements can
be evaluated in a similar fashion. It follows from the second term of Eq.\eqref{eq 3.6},
the matrix elements from the photon self-interaction contribution,
\begin{eqnarray}
\rho _{self-int}^{\left( 1\right) }\left( k^{2}\right) \tau \left(
k_{0}\right) &=&\int dp_{1}dp_{2}\int d^{2}zd^{2}w\delta ^{\left( 2\right) }\left(
k-p_{1}-p_{2}\right) \eta _{\mu \nu }\Delta ^{\mu \sigma (ret)}\left(
-z\right) \Delta ^{\nu \rho (ret)}\left( -w\right)  \notag \\
&& \times\underset{i,j}{\sum }\left\langle \Omega \left\vert  j^{in}
_{\sigma }\left( z\right) \right\vert p_{1},i;p_{2},j\right\rangle
\left\langle p_{1},i;p_{2},j\left\vert  j^{in}_{\rho }\left(
w\right) \right\vert \Omega \right\rangle ,
\label{eq 3.9} \\
&=&\frac{g^{2}}{4\pi }\frac{1}{\left( k^{2}\right) ^{2}}\int d^{2}p\tau
\left( p_{0}\right) \delta \left( p^{2}\right) \tau \left(
k_{0}-p_{0}\right) \delta \left( \left( k-p\right) ^{2}\right)\notag \\
&&\times \left[ \left( p\circ p\right)-\left( k\circ k\right) -\left( k\circ
p\right)  +\frac{6}{k^{2}}\left( k\times
p\right) ^{2}\right] , 
\end{eqnarray}%
where $\left( a\circ b\right) =a_{\alpha }\theta ^{\alpha \lambda }\theta
_{\lambda \sigma }b^{\sigma }$ and $\left( a\times b\right) =a_{\alpha
}\theta ^{\alpha \lambda }b_{\lambda }$. Nevertheless, from a
straightforward computation of the remaining matrix elements, Eq. \eqref{eq 3.6} is, thus, written as
\begin{eqnarray}
\rho ^{\left( 1\right) }\left( k^{2}\right) \tau \left( k_{0}\right) &=&
\frac{g^{2}}{4\pi \left( k^{2}\right) ^{2}}\bigg\{8m^{2}\left[ 1+%
\frac{\left( k\circ k\right) }{16}\left( k^{2}-4m^{2}\right) \right]\notag \\
&&\times \int d^{2}p\tau \left( p_{0}\right) \tau \left( k_{0}-p_{0}\right)
\delta \left( p^{2}-m^{2}\right) \delta \left( \left( k-p\right)
^{2}-m^{2}\right) \notag\\
&&+\int d^{2}p\tau \left( p_{0}\right) \delta \left( p^{2}\right) \tau
\left( k_{0}-p_{0}\right) \delta \left( \left( k-p\right) ^{2}\right) \notag \\
&&\times  \left[ \left( p\circ p\right)-\left( k\circ k\right) -\left( k\circ
p\right) +\frac{6}{k^{2}}\left( k\times
p\right) ^{2}\right] \bigg\}. \label{eq 3.11}
\end{eqnarray}%
Evaluating the momentum integral, we find the result for the propagator
\begin{eqnarray}
i\mathcal{D}\left( k^{2}\right) &=&\frac{2}{k^{2}}+\frac{g^{2}m^{2}}{\pi k^{2} }%
\int_{4m^{2}}^{\infty }\frac{d\chi }{\chi ^{2}\left( k^{2}-\chi -i\epsilon
\right) }\frac{\left[ 1+\frac{\left( k\circ k\right) }{16}\left( \chi
-4m^{2}\right) \right] }{\sqrt{1-\frac{4m^{2}}{\chi }}}  +\frac{g^{2}\theta ^{2}}{4\pi k^{2}}\int_{\lambda ^{2}}^{\infty }\frac{d\chi }{%
\chi \left( k^{2}-\chi -i\epsilon \right) },  \label{eq 3.13}
\end{eqnarray}%
finally yielding the expression
\begin{eqnarray}
i\mathcal{D}\left( k^{2}\right) &=&\frac{2}{k^{2}}
+\frac{g^{2}}{4\pi }\frac{\theta ^{2}}{k^{4}}\ln \left[ 1-\frac{k^{2}}{%
\lambda ^{2}}\right] \notag \\
&& +\frac{g^{2}}{8\pi k^{4}}
\left[ \left( 4+k^{2}\theta ^{2}m^{2}\right) -\frac{m^{2}\left(
16-k^{2}\theta ^{2}\left( k^{2}-4m^{2}\right) \right) }{\sqrt{k^{2}\left(
4m^{2}-k^{2}\right) }}\csc ^{-1}\left[ \frac{2m}{\sqrt{k^{2}}}\right] \right].  \label{eq 3.14}
\end{eqnarray}%
It is the first term inside the brackets of Eq.\eqref{eq 3.14} that, when the
limit $m\rightarrow 0$ is taken and the vacuum polarization bubbles are
summed, gives rise to the well-known Schwinger mass $m_{A}^{2}=\frac{g^{2}}{%
\pi }$ of the photon \cite{2}. In order to obtain a simplified expression of Eq.\eqref{eq 3.14} and to compare with a previous result \cite{4a}, we have
made use of the fact that in a two-dimensional spacetime, the noncommutative
matrix $\theta ^{\mu \nu }$ can be expressed as $\theta ^{\mu \nu }=\theta
\epsilon ^{\mu \nu }$, where $\epsilon ^{\mu \nu }$ is the two-dimensional
Levi-Civit\`{a} tensor. Therefore, with this particular choice, we have
obtained that the self-interaction contribution is%
\begin{equation}
i\mathcal{D}_{self-int}^{\left( 1\right) }\left( k^{2}\right) =\frac{g^{2}}{4\pi }%
\frac{\theta ^{2}}{k^{4}}\ln \left[ 1-\frac{k^{2}}{\lambda ^{2}}\right] .
\label{eq 3.10}
\end{equation}%
This result is in contrast with the one obtained previously in Ref.\cite{4a},
where it was discussed, through one-loop diagram evaluation, that this contribution gives rise to a higher-order
term (when summed into the complete propagator) dynamically
generated by quantum corrections and that it is ultraviolet finite. Moreover,
in Ref.\cite{4a}, the propagator behavior in the infrared sector was not clear,
which becomes clearly important and transparent in the framework of dispersion
integrals. Furthermore, in order to make the integral infrared finite, we had
to introduce a finite $\lambda $ photon mass in Eq.\eqref{eq 3.13} $\tau \left(
k^{2}\right) \rightarrow \tau \left( k^{2}-\lambda ^{2}\right) $, therefore, showing
that this term is, in fact, a purely infrared effect, as it is the mechanism behind the photon mass generation and that an analysis on that
plays an important role in the correct interpretation of this term \cite{36}. However,
it is clear from the expression \eqref{eq 3.10} that this contribution is
not actually a higher-order term in any plausible limit when the smallness of $\lambda$ is taken into
account.


\subsection{($2+1 $)-dimensional case}

In the hope of learning useful things about the intriguing well-recognized four-dimensional problems,
a lot of attention has been paid to the analysis of general properties of the simple setting of three-dimensional field theories over the years \cite{20},
in particular $QED_3$. For instance, Ref.\cite{21} provided an unambiguous
answer to the question about whether the dynamically generated photon mass is different from zero
\cite{5}. Now we revisit this issue in light of Ref.\cite{21}
but for a noncommutative theory looking at whether a noncommutative contribution is present. The calculation of the spectral density function on the $QED_3$
follows the same lines as presented above for the two-dimensional case. However, before starting
the calculation, let us recall important points about the general
structure of $\rho _{\mu \nu }$. It follows from the gauge and Lorentz
invariance that $\rho _{\mu \nu }$ can be expressed as%
\begin{equation}
\rho _{\mu \nu }\left( k^{2}\right) =\left( \eta _{\mu \nu }-\frac{k^{\mu
}k^{\nu }}{k^{2}}\right) \rho _{S}\left( k^{2}\right) +i\epsilon _{\mu \nu
\sigma }k^{\sigma }\rho _{A}\left( k^{2}\right) ,  \label{eq 4.0}
\end{equation}%
where it was added the scalar functions $\rho _{i}$ to the symmetric and
antisymmetric sectors. Furthermore, these functions are determined as
follows:%
\begin{eqnarray}
\rho _{S}\left( k^{2}\right)& =&\frac{1}{2}\rho _{\mu }^{\mu }\left(
k^{2}\right) , \label{eq 4.1a} \\
\rho _{A}\left( k^{2}\right) &=&-\frac{i}{2k^{2}}k_{\lambda }\epsilon ^{\mu
\nu \lambda }\rho _{\mu \nu }\left( k^{2}\right) .  \label{eq 4.1b}
\end{eqnarray}%
For instance, for the one-particle contribution, one obtains from
Eqs.\eqref{eq 2.2} and \eqref{eq 4.1a} the following result for the symmetric form factor: $
\rho _{S}^{\left( 0\right) }\left( k^{2}\right) =\frac{3}{2}\delta \left(
k^{2}-\mu^{2}\right) $. Now, for the two-particle contribution, we make use again of the same
arguments presented above to construct the intermediate state, which led to the expression \eqref{eq 3.6}. Therefore, by the same arguments, it
follows that the spectral density function is given by
\begin{eqnarray}
\frac{\rho ^{\left( 1\right)\mu \nu  }\left( k^{2}\right) \tau \left(
k_{0}\right) }{\left( 2\pi \right) ^{2}} &=&\int d^{2}p_{1}d^{2}p_{2}\int
d^{3}zd^{3}w\delta ^{\left( 3\right) }\left( k-p_{1}-p_{2}\right) \Delta
^{\mu \sigma (ret)}\left( -z\right) \Delta ^{\nu \rho (ret)}\left(
-w\right)  \notag \\
&&\times \bigg\{\underset{r,s}{\sum }\left\langle \Omega \left\vert
j^{in}_{\sigma }\left( z\right) \right\vert
p_{1},r;p_{2},s\right\rangle \left\langle p_{1},r;p_{2},s\left\vert
j^{in}_{\rho }\left( w\right) \right\vert \Omega \right\rangle \notag \\
&&+\underset{i,j}{\sum }\left\langle \Omega \left\vert  j^{in}
_{\sigma }\left( z\right) \right\vert p_{1},i;p_{2},j\right\rangle
\left\langle p_{1},i;p_{2},j\left\vert  j^{in}_{\rho }\left(
w\right) \right\vert \Omega \right\rangle \bigg\}. \label{eq 4.3}
\end{eqnarray}%
Where $j$ is given by Eq.\eqref{eq 1.15} and the retarded Green's function by %
Eq.\eqref{eq 1.17}. Next, by a straightforward but rather lengthy calculation on the
matrix elements, one can obtain the following expression for the symmetric
two-particle contribution \eqref{eq 4.1a}:
\begin{eqnarray}
\rho _{S}^{\left( 1\right) }\left( k^{2}\right) \tau \left( k_{0}\right) &=&%
\frac{g^{2}}{2\left( 2\pi \right) ^{2}}\frac{1}{\left( k^{2}-\mu^{2}\right) }%
\bigg\{\left[ \left( 1+\frac{4m^{2}}{k^{2}}\right) +\frac{m^{2}}{4}\left(
k\circ k\right) \left( 1-\frac{4m^{2}}{k^{2}}\right) \right] \notag
\\
&&\times \int d^{3}p\tau \left( p_{0}\right) \tau \left( k_{0}-p_{0}\right)
\delta \left( p^{2}-m^{2}\right) \delta \left( \left( k-p\right)
^{2}-m^{2}\right)  \notag \\
&&-\int d^{3}p\tau \left( p_{0}\right) \tau \left( k_{0}-p_{0}\right) \delta
\left( p^{2}\right) \delta \left( \left( k-p\right) ^{2}\right) \left[
3\left( k\times p\right) ^{2}-k^{2}\left( k\circ k\right) \right] \bigg\},
 \label{eq 4.4}
\end{eqnarray}%
whereas for the antisymmetric contribution \eqref{eq 4.1b}, one finds
\begin{eqnarray}
\rho _{A}^{\left( 1\right) }\left( k^{2}\right) \tau \left( k_{0}\right) &=&-%
\frac{ig^{2}}{\left( 2\pi \right) ^{2}}\frac{1}{k^{2}k^{2}\left(
k^{2}-\mu^{2}\right) }\int d^{3}p\tau \left( p_{0}\right) \tau \left(
k_{0}-p_{0}\right) \delta \left( p^{2}-m^{2}\right) \delta \left( \left(
k-p\right) ^{2}-m^{2}\right) \notag \\
&&\times \left[ 2im^{2}\epsilon _{\lambda \sigma \pi }p^{\lambda }\tilde{k}%
^{\sigma }k^{\pi }+mk^{2}\left( 2i-\left( k\times p\right) \right) \right] .  \label{eq 4.5}
\end{eqnarray}%
Therefore, it follows, by solving the remaining momentum integration, the
explicit expressions%
\begin{eqnarray}
\rho _{S}^{\left( 1\right) }\left( k^{2}\right) &=&\frac{\alpha }{4}\frac{1}{%
\sqrt{k^{2}}\left( k^{2}-\mu^{2}\right) }\bigg\{\frac{11 k^{2}}{8}\left( k\circ
k\right) \tau \left(k^{2}\right)  \notag \\
&&+\left[ \left( 1+\frac{4m^{2}}{k^{2}}\right) +\frac{m^{2}}{4}\left( k\circ
k\right) \left( 1-\frac{4m^{2}}{k^{2}}\right) \right] \tau \left(
k^{2}-4m^{2}\right) \bigg\},  \label{eq 4.6}
\end{eqnarray}
and
\begin{equation}
\rho _{A}^{\left( 1\right) }\left( k^{2}\right) =\frac{\alpha }{k^{2}\left(
k^{2}-\mu^{2}\right) }\frac{m}{\sqrt{k^{2}}}\tau \left( k^{2}-4m^{2}\right) ,
\label{eq 4.7}
\end{equation}%
where $\alpha =\frac{g^{2}}{4\pi }$. As said above, it is a well-known feature that in
three dimensions the photon field acquires a non-null mass \cite{21}; hence, we are interested in analyzing here
whether this mass of photon is changed due to NC effects. By
means of that, we consider the proper vacuum polarization insertions,
\begin{equation}
\left( \mathcal{D}^{-1}\right) _{\mu \nu }=\left( D^{-1}\right) _{\mu \nu
}-i\Pi _{\mu \nu },  \label{eq 4.30}
\end{equation}%
where the free propagator in a general gauge parameter $\xi $ is%
\begin{equation}
\left( D^{-1}\right) _{\mu \nu }=ik^{2}\left( \eta _{\mu \nu }-\frac{k_{\mu
}k_{\nu }}{k^{2}}+\frac{i\mu}{k^{2}}\epsilon _{\mu \nu \lambda }k^{\lambda }+%
\frac{1}{\xi }\frac{k_{\mu }k_{\nu }}{k^{2}}\right) ,
\end{equation}%
whereas the vacuum polarization tensor is
\begin{equation}
\Pi _{\mu \nu }\left( k\right) =\left( \eta _{\mu \nu }-\frac{k_{\mu }k_{\nu
}}{k^{2}}\right) \Pi _{S}\left( k^{2}\right) +i\epsilon _{\mu \nu \sigma
}k^{\sigma }\Pi _{A}\left( k^{2}\right) .
\end{equation}%
Furthermore, the scalar polarization functions $\Pi _{S}$ and $\Pi _{A}$ are related with the scalar
spectral functions $\rho _{S}$ and $\rho _{A}$ by means of the relation
\begin{equation}
\Pi _{\mu \nu }\left( k\right) =\int_{0}^{\infty }d\chi \sigma _{\mu \nu
}\left( \chi \right) \frac{1}{k^{2}-\chi -i\epsilon },  \label{eq 4.31}
\end{equation}%
where the spectral function $\sigma _{\mu\nu}$ includes contributions of all loops to
$\rho _{\mu\nu}$. In order to compute $\mathcal{D}$ from Eq.\eqref{eq 4.30}, one can make use of
the following set of orthogonal projection operators%
\begin{eqnarray}
P_{\mu \nu }^{\left( 1\right) } &=&\frac{1}{2}\left( \eta _{\mu \nu }-\frac{%
k_{\mu }k_{\nu }}{k^{2}}+i\epsilon _{\mu \nu \lambda }\frac{k^{\lambda }}{%
\sqrt{k^{2}}}\right) ,  \notag \\
P_{\mu \nu }^{\left( 2\right) } &=&\frac{1}{2}\left( \eta _{\mu \nu }-\frac{%
k_{\mu }k_{\nu }}{k^{2}}-i\epsilon _{\mu \nu \lambda }\frac{k^{\lambda }}{%
\sqrt{k^{2}}}\right) ,  \label{eq 4.32} \\
P_{\mu \nu }^{\left( 3\right) } &=&\frac{k_{\mu }k_{\nu }}{k^{2}},  \notag
\end{eqnarray}%
to, therefore, obtain%
\begin{eqnarray}
\mathcal{D}^{-1} &=&i\left( k^{2}\left( 1+\frac{\mu}{\sqrt{k^{2}}}\right) -\Pi
_{S}-\sqrt{k^{2}}\Pi _{A}\right) P^{\left( 1\right) }  \label{eq 4.33}\\
&&+i\left( k^{2}\left( 1-\frac{\mu}{\sqrt{k^{2}}}\right) -\Pi _{S}+\sqrt{k^{2}}%
\Pi _{A}\right) P^{\left( 2\right) }+\frac{i}{\xi }k^{2}P^{\left(
3\right) }.  \notag
\end{eqnarray}%
Finally, after some algebraic manipulation, one can find%
\begin{equation}
i\mathcal{D}_{\lambda \nu }=\frac{1}{k^{2}-\Pi \left( k^{2}\right) }\left[ \eta
_{\lambda \nu }-\frac{k_{\lambda }k_{\nu }}{k^{2}}-\frac{i}{k^{2}}\epsilon _{\lambda \nu
\sigma }k^{\sigma }\frac{\mu-\Pi _{A}}{1-\frac{\Pi _{S}}{k^{2}}}\right],  \label{eq 4.34}
\end{equation}%
for $\xi=0$, where we have defined%
\begin{equation}
\Pi \left( k^{2}\right) =\Pi _{S}+\frac{\left( \mu-\Pi _{A}\right) ^{2}}{1-%
\frac{\Pi _{S}}{k^{2}}}.  \label{eq 4.35}
\end{equation}%
It follows that, in the second-order perturbation theory, the following
expression for the photon mass%
\begin{equation}
\Pi^{\left( 1\right) } \left( 0\right) =\Pi^{\left( 1\right) } _{S}\left(
0\right) +\frac{\left( \mu-\Pi^{\left( 1\right) } _{A}\left( 0\right) \right)
^{2}}{1-\frac{\Pi^{\left( 1\right) } _{S}\left( 0\right) }{k^{2}}}.
\label{eq 4.36}
\end{equation}%
Now, by calculating the functions $\Pi _{S}^{\left( 1\right) }$ and $\Pi _{A}^{\left(
1\right) }$ through Eqs.\eqref{eq 4.31}, \eqref{eq 4.6} and \eqref{eq 4.7} evaluating the integration in $\chi$ for the region of interest $%
k^{2}<4m^{2}$, and, at last, expanding the expressions to $k^{2}\rightarrow 0$%
, we find%
\begin{eqnarray}
\underset{k^{2}\rightarrow 0}{\lim }\frac{\Pi _{S}^{\left( 1\right) }\left(
k^{2}\right) }{k^{2}}&=&\frac{\alpha }{8\mu^{3}}\bigg\{ 2m\mu\left(4- m^{2}\left(
k\circ k\right)\right)  \label{eq 4.37} \\
&&+\bigg[ \left( k\circ k\right) m^{2}\left(
4m^{2}-\mu^{2}\right) -4\left( \mu^{2}+4m^{2}\right) \bigg] \coth ^{-1}\left[
\frac{2m}{\mu}\right] \bigg\} , \notag
\end{eqnarray}%
and%
\begin{equation}
\underset{k^{2}\rightarrow 0}{\lim }\Pi _{A}^{\left( 1\right) }\left(
k^{2}\right) =-\frac{g^{2}}{4\pi }\frac{2m}{\mu}\coth ^{-1}\left[ \frac{2m}{\mu}%
\right] .  \label{eq 4.38}
\end{equation}%
In order to trace a parallel with the known result for $QED$ \cite{21}, we take the limit $%
\mu\rightarrow 0$ in the above expressions \eqref{eq 4.37} and \eqref{eq 4.38}. It follows, therefore, a non-null mass for the photon field
\begin{equation}
\Pi \left( 0\right) =\frac{\alpha ^{2}}{1+\frac{\alpha }{12m}\left[
8+m^{2}\left( k\circ k\right) \right] }\neq 0. \label{eq 4.38a}
\end{equation}%
Such expression reproduces the known result from $QED_{3}$ for $k_{nc}^{2}=\left(
k\circ k\right) \rightarrow 0$. In $QED_3$ there is a proof in Ref.\cite{27} where
it was shown that all contributions to the mass
from other graphs vanish identically; therefore, it is plausible to say for $QED_{3}$
that a nonperturbative mass is dynamically generated. However,
there is no known proof (or discussion) for the noncommutative $QED_3$ counterpart,
which makes it impossible for us to say anything about whether or not the above result \eqref{eq 4.38a} is nonperturbative in nature.

\section{Concluding Remarks}


\label{sec:4}

This paper presents a study of the two- and three-dimensional NC quantum
electrodynamics in light of the K\"{a}ll\'{e}n-Lehmann spectral representation.
Our main interest here is studying NC extensions of well-known
quantum field theories to look for NC effects on its deviations. These two
models $QED_{2}$ and $QED_{3}$ were extensively studied in several areas
of theoretical physics and have long been recognized as laboratories of
testing new ideas on a simpler setting, especially on condensed matter and statistical systems, for instance, the quantum Hall effect.

We begin by discussing NC quantum electrodynamics defined on an $\omega $%
-dimensional spacetime. After discussing some of the symmetries present on the action, we
make use of the Seiberg-Witten map to determine an action. Interesting features
are present on that; for instance, it is due to the noncommutativity of spacetime
that there are nonlinear terms on the field strength. Furthermore, for $%
\omega >2$, this action also provides a suitable framework for studies on
Lorentz-violating extension. Next, restricting our attention to the cases of
two and three dimensions, it is possible to obtain two Lagrangian functions, from
which we derive the respective equations of motion for the gauge and
fermionic fields. Furthermore, based on the Yang-Feldman-K\"{a}ll\'{e}n formalism, it is
possible to find solutions to the Heisenberg operator equations of motion
solutions that are an important ingredient of the K\"{a}ll\'{e}n-Lehmann
representation.

On the K\"{a}ll\'{e}n-Lehmann representation, we first introduce the
spectral density function for the gauge field followed by a discussion on its general
properties and present, finally, its relation with the exact Feynman
propagators as well. The main aim of the present paper is to evaluate the
one- and two-particle contribution to the gauge field spectral
function in two and three dimensions. Based on that, for
the photon field, after calculating the two-particle contribution ($%
g^{2}$ order), we obtain in two dimensions for the self-interaction
contribution an interesting result which differs from a previous one \cite{4a}, where
it states that this same contribution gives rise to a higher-derivative term.
The main difference we show here is that this term depends explicitly on the infrared
parameter, a photon mass $\lambda$, and in any plausible limit taking into account the
smallness of $\lambda$, this term does not generate a higher-order term as stated in Ref.\cite{4a}. This consists of an analysis
that was not described earlier and clearly plays an important role in order to
obtain a consistent interpretation and correct expression arising from the infrared
sector, and it is automatically fulfilled in the dispersion relation
calculation. Furthermore, based on the well-known result in
three dimensions, where a dynamical mass for the photon is generated from
radiative calculation, we derive the relation between the self-energy
function with the propagator pole in order to see whether and how the noncommutativity
affects the value of the photon mass.

For different reasons, much interest is still present in studying lower-dimensional
field theory in the most different quantization process, since they always
provide a rich testing ground to study the most diverse ideas of more
realistic systems, especially in the well-recognized four-dimensional problems. We still have interest in studying some thermodynamical
features of these systems \cite{26}, in particular, three-dimensional NCQED. Moreover, nonrelativistic field theories
around Lifshitz points \cite{34} with anisotropic scaling between space and time have come to our attention and certainly deserve
to be discussed in lower-dimensional theories carefully and in further details \cite{33}.
These issues and others will be further elaborated, investigated, and
reported elsewhere.


\subsection*{Acknowledgments}

The authors would like to thank the referee for his/her
comments and suggestions to improve this paper. R.B. thanks FAPESP for full support, T.R.C. thanks CAPES for full support and B.M.P.
thanks CNPq and CAPES for partial support.


\appendix


\section{Notation and Identities}

\label{app:A}

In this appendix, we fix our notation and review some useful identities for
two- and three-dimensional spacetime. First, for two dimensions, our
convention for the metric and the $\gamma $-matrices representation are%
\begin{equation*}
\eta _{\mu \nu }=\left(
\begin{array}{cc}
1 & 0 \\
0 & -1%
\end{array}%
\right) ,
\end{equation*}%
and%
\begin{equation*}
\gamma _{0}=\sigma _{1},\quad \gamma _{1}=-i\sigma _{2},\quad \gamma
_{5}=\gamma _{0}\gamma _{1}=\sigma _{3},
\end{equation*}%
where $\left\{ \sigma _{i}\right\} $ are the Pauli matrices. We have%
\begin{equation}
\left[ \gamma _{\mu },\gamma _{\nu }\right] _{+}=2\eta _{\mu \nu },\quad %
\left[ \gamma _{\mu },\gamma _{\nu }\right] =-2\epsilon _{\mu \nu }\gamma
_{5},  \label{eq a.1}
\end{equation}%
from which it follows $\gamma _{\mu }\gamma _{5}=\epsilon _{\mu \nu }\gamma
^{\nu }$ and $\epsilon _{\mu \nu }\epsilon ^{\alpha \beta }=\delta _{\mu
}^{\alpha }\delta _{\nu }^{\beta }-\delta _{\mu }^{\beta }\delta _{\nu
}^{\alpha }$. Moreover, for the three-dimensional case, we have%
\begin{equation*}
\eta _{\mu \nu }=\left(
\begin{array}{ccc}
1 & 0 & 0 \\
0 & -1 & 0 \\
0 & 0 & -1%
\end{array}%
\right) ,
\end{equation*}%
and%
\begin{equation*}
\gamma _{0}=\sigma _{3},\quad \gamma _{1}=i\sigma _{1},\quad \gamma
_{2}=i\sigma _{2}.
\end{equation*}%
We also have the algebra%
\begin{equation}
\left[ \gamma _{\mu },\gamma _{\nu }\right] _{+}=2\eta _{\mu \nu },\quad %
\left[ \gamma _{\mu },\gamma _{\nu }\right] =-2i\epsilon _{\mu \nu \lambda
}\gamma ^{\lambda },  \label{eq a.2}
\end{equation}%
and some relevant properties%
\begin{equation}
\gamma _{\mu }\gamma ^{\mu }=3,\quad \epsilon _{\alpha \beta \lambda
}\epsilon ^{\alpha \beta \lambda }=3!,  \label{eq a.3}
\end{equation}%
\begin{equation}
\text{Tr}\left( \gamma _{\mu }\right) =0,\quad \text{Tr}\left( \gamma _{\mu
}\gamma _{\nu }\right) =2\eta _{\mu \nu },\quad \text{Tr}\left( \gamma _{\mu
}\gamma _{\nu }\gamma _{\sigma }\right) =-2i\epsilon _{\mu \nu \sigma }.
\label{eq a.4}
\end{equation}


\section{Integral calculation}

\label{app:B}

In this appendix, we provide a calculation of some relevant
integrals. We have here three types of dispersion integrals: a scalar, vector, and
tensor type. First, the two-dimensional integrals,
\begin{eqnarray}
I^{\left( 1\right) }\left( k\right) =\int d^{2}p\tau \left( p_{0}\right)
\tau \left( k_{0}-p_{0}\right) \delta \left( p^{2}-m^{2}\right) \delta
\left( \left( k-p\right) ^{2}-m^{2}\right) ,  \label{eq b.1}
\end{eqnarray}%
for a timelike $k$, we can go to a Lorentz frame where $k=\left(
k_{0},0\right) $. In this frame follows%
\begin{equation}
I^{\left( 1\right) }\left( k_{0}\right) =\int \frac{dp}{2E_{p}}\tau
\left( k_{0}-E_{p}\right) \delta \left( k_{0}^{2}-2k_{0}E_{p}\right) ,
\end{equation}%
where%
\begin{equation*}
E_{p}=\sqrt{\mathbf{p}^{2}+m^{2}}=\frac{k_{0}}{2},
\end{equation*}%
which implies $\left\vert \mathbf{p}\right\vert =\sqrt{k_{0}^{2}/4-m^{2}}$.
Therefore, returning to an arbitrary Lorentz frame, it follows that%
\begin{equation}
I^{\left( 1\right) }\left( k\right) =\frac{1}{4k^{2}}\frac{1}{\sqrt{1-%
\frac{4m^{2}}{k^{2}}}}\tau \left( k^{2}-4m^{2}\right) \tau \left(
k_{0}\right) .  \label{eq b.2}
\end{equation}%
Another relevant two-dimensional integral is%
\begin{equation}
I^{\left( 2\right) }\left( p\right) =\int d^{2}k\tau \left( k_{0}\right)
\tau \left( p_{0}-k_{0}\right) \delta \left( k^{2}-m^{2}\right) \delta
\left( \left( p-k\right) ^{2}\right) ,  \label{eq b.3}
\end{equation}%
and%
\begin{equation*}
I^{\left( 2\right) }\left( p\right) =\int \frac{dk}{2E_{k}}\tau \left( p_{0}-E_{k}\right) \delta \left(
p_{0}^{2}-2p_{0}E_{k}+m^{2}\right) ,
\end{equation*}%
where we had taken again a timelike $p$ in the form $p=\left(
p_{0},0\right) $, and from the above expression follows $\left\vert \mathbf{%
k}\right\vert =\frac{p_{0}^{2}-m^{2}}{2p_{0}}$ and for $p_{0}>0$, $%
p_{0}^{2}-m^{2}>0$. Therefore%
\begin{eqnarray*}
I^{\left( 2\right) }\left( p\right) =\frac{1}{4\left\vert p_{0}\right\vert }\tau \left(
p_{0}^{2}-m^{2}\right) \tau \left( p_{0}\right) \int dE_{k} \frac{\delta \left( \frac{%
p_{0}^{2}+m^{2}}{2p_{0}}-E_{k}\right)}{\sqrt{%
E_{k}^{2}-m^{2}}} .
\end{eqnarray*}%
At last,%
\begin{equation}
I^{\left( 2\right) }\left( p\right) =\frac{1}{2}\frac{1}{\sqrt{\left(
p^{2}-m^{2}\right) ^{2}}}\tau \left( p^{2}-m^{2}\right) \tau \left(
p_{0}\right) .  \label{eq b.4}
\end{equation}
Moreover, for three dimensions, the scalar integrals are obtained following
the same steps as before. We evaluate, thus, a vector and tensor integral for
three dimensions. First,%
\begin{eqnarray}
I_{\mu }^{\left( 3\right) }(k)=\int d^{3}p\tau \left( p_{0}\right)  \tau \left( k_{0}-p_{0}\right) \delta\left( p^{2}-m^{2}\right)\delta \left(
\left( k-p\right) ^{2}-m^{2}\right) p_{\mu };  \label{eq b.5}
\end{eqnarray}%
and it follows for $\mu \neq 0$ that $I_{\mu }^{\left( 3\right) }=0$ by
symmetry. Thus, we have%
\begin{equation}
I_{\mu }^{\left( 3\right) }(k)=\frac{\pi  k_{\mu }  }{4\sqrt{k^{2}}}\tau \left(
k^{2}-4m^{2}\right) \tau \left( k_{0}\right).  \label{eq b.6}
\end{equation}%
It also follows, without further complication, that%
\begin{eqnarray}
I_{\mu }^{\left( 4\right) } (p)&=&\int d^{3}k\tau \left( k_{0}\right)  \tau \left( p_{0}-k_{0}\right)\delta \left( k^{2}-m^{2}\right)\delta \left(
\left( p-k\right) ^{2}\right) k_{\mu },   \notag \\
&=&\frac{\pi p_{\mu }}{4\sqrt{p^{2}}}\left( 1+\frac{m^{2}}{p^{2}}\right) \tau \left(
p^{2}-m^{2}\right) \tau \left( p_{0}\right) . \label{eq b.7}
\end{eqnarray}%
Now, the tensor integral,%
\begin{eqnarray}
I_{\mu \nu }^{\left( 5\right) }(p) &=&\int d^{3}q\tau \left( q_{0}\right)
 \tau \left( p_{0}-q_{0}\right)\delta \left( q^{2}\right)\delta \left(
\left( p-q\right) ^{2}\right) q_{\mu }q_{\nu }   \notag \\
&=& \left( -\eta _{\mu \nu }+3\frac{p_{\mu }p_{\nu }}{p^{2}}\right)\frac{\pi \sqrt{p^{2}}}{16}\tau \left(
p^{2}\right) \tau \left( p_{0}\right) . \label{eq b.8}
\end{eqnarray}%
As remarked in the beginning of this appendix, we listed and
evaluated explicitly here some relevant dispersion integrals. Moreover, it is worth of saying that
the remaining integrals are simpler and follow the same lines.


\end{document}